\documentclass[preprint,12pt]{elsarticle}
\usepackage{graphicx}

\journal{Elsevier}  

\begin{document}
\begin{frontmatter}

\title{Near-threshold $K^- d$ scattering and properties of
kaonic deuterium}

\author{N.V. Shevchenko}
\address{Nuclear Physics Institute, 25068 \v{R}e\v{z}, Czech Republic}

\date{\today}

\begin{abstract}
We calculated the $1s$ level shifts and widths of kaonic deuterium,
corresponding to accurate results on near-threshold
antikaon - deuteron scattering. The Lippmann-Schwinger
eigenvalue equation with a strong $K^- - d$ and Coulomb potentials was
solved. The two-body $K^- - d$ potentials reproduce the near-threshold
elastic amplitudes of $K^- d$ scattering obtained from the three-body
Alt-Grass\-berger-Sand\-has equations with the coupled channels
using four versions of the $\bar{K}N - \pi \Sigma$ potentials.
Both new $\bar{K}N - \pi \Sigma$ potentials reproducing 
the very recent SIDDHARTA data on kaonic hydrogen and
our older potentials reproducing KEK data have one- or two-pole
versions of the $\Lambda(1405)$ resonance and reproduce 
experimental data on $K^- p$ scattering.
\end{abstract}


\end{frontmatter}

\section{Introduction}
\label{intro.sect}

Kaonic atoms as well as scattering of antikaons on nuclei can be
used for investigation of antikaon-nucleon interaction. Deuteron is
a particularly promising target since dynamics of the
few-body $K^- d$ system can be properly described by Faddeev equations.
In our previous paper~\cite{my_Kd} we used the equations for
calculations of the $K^- d$ scattering length $a_{K^- d}$ and investigated
its dependence on the two-body input. In particular, we were interested in
dependence of $a_{K^- d}$ on phenomenological models of the $\bar{K}N$
interaction with one or two poles forming $\Lambda(1405)$ resonance
since the question of number of the poles is quite actual
(see e.g.~\cite{ECT2009,ECT2010}).

In spite of the fact that the most of the ``chirally based'' models
of the interaction have two poles, such two-pole structure and
energy dependence are not necessary for reproducing experimental
data on $K^- p$ scattering and kaonic hydrogen.
It was clearly shown in~\cite{my_Kd,ourPRC_isobreak}, where all
the data were described by energy-independent $\bar{K}N - \pi \Sigma$
potentials having one- or two-pole structure with equally high
accuracy. It was also demonstrated that no experimental data
prove the two-pole structure of the $\Lambda(1405)$ resonance.

However, the scattering length, in contrast to the $1s$ level shift
and width of kaonic deuterium, cannot be measured directly. The
task of calculation of the kaonic deuterium characteristics is quite
actual in view of SIDD\-HAR\-TA-2 experiment~\cite{SIDDHARTA-2}, planning
measurements of the atom.

Mixture of strong and Coulomb interactions complicates theoretical
investigation of low levels of hadronic atoms using methods of few-body
physics. Due to this we calculated the energy of the $1s$ level
considering kaonic deuterium as a two-body $K^- - d$ system.
For this purpose a strong $K^- - d$ potential,
reproducing the $K^- d$ scattering observables obtained in the advanced
calculation, was constructed. However, additional three-body calculations
were necessary to be done before. 

First, the potentials
used in the calculations in~\cite{my_Kd} reproduce the KEK data on the
kaonic hydrogen characteristics together with all experimental data
on low-energy $K^- p$ scattering. Keeping in mind the very recent data on
kaonic hydrogen measured in the SIDDHARTA experiment~\cite{SIDDHARTA1s},
we constructed the new $\bar{K}N - \pi \Sigma$
potentials, which reproduce the newest experimental data, and repeated
the calculations of the $K^- d$ scattering length.

Second, knowledge of the scattering length is not enough for construction
of the $K^- - d$ potential, therefore, we additionally calculated the
near-threshold $K^- d$  scattering amplitudes for all four models of
antikaon-nucleon interaction: used in~\cite{my_Kd} and the new ones.
Effective ranges were calculated as well. 
Finally, the results of the three-body calculations were used for
construction of the strong $K^- - d$ potentials. The $1s$ level shifts
and widths of kaonic deuterium were then obtained by solving the
Lippmann-Schwinger eigenvalue equation with the constructed strong
potentials and directly included Coulomb interaction.

An effective two-body $K^- - d$ potential was used for calculation
of level shifts and widths of kaonic deuterium in~\cite{BarretDeloff}.
The zero-range potential, used in the paper, reproduces a three-body
$K^- d$ scattering length only.

The first subsection of the next section is devoted to the new
$\bar{K}N - \pi \Sigma$ potentials, while the rest of the two-body input
and the three-body formalism used for the calculations of the
near-threshold elastic $K^- d$ amplitudes are described in
subsection~\ref{formalism.sect}.
The obtained $K^- d$ scattering lengths and three-body amplitudes
are demonstrated and discussed in the last subsection of
Section~\ref{Kdscattering.sec}. Section \ref{deuterium.sec}
contains description of the two-body strong $K^- - d$
potential together with the method of the calculation
of the $1s$ level energy of kaonic deuterium.
The results are shown and discussed in section \ref{results.sec},
the last section concludes the article.

\section{Near-threshold $K^- d$ scattering}
\label{Kdscattering.sec}

\subsection{$\bar{K}N - \pi \Sigma$ potentials reproducing SIDDHARTA data}
\label{SIDDpotentials.sect}

All $\bar{K}N - \pi \Sigma$ potentials used in
our previous work~\cite{my_Kd} for the $K^- d$
scattering length calculations reproduce,
among other experimental data, the $1s$ level shift $\Delta E_{1s}$
and width $\Gamma_{1s}$ of kaonic hydrogen, measured in
KEK~\cite{KEK1s}:
\begin{equation}
\label{KEK}
\Delta E_{1s}^{\rm KEK} = -323 \pm 63 \pm 11 \; {\rm eV}, \quad
\Gamma_{1s}^{\rm KEK} = 407 \pm 208 \pm 100 \; {\rm eV.} 
\end{equation}
In what follows the two representative
potentials from Ref.~\cite{my_Kd}, having one and two poles forming
the $\Lambda(1405)$ resonance, will be denoted by
$V^{\rm 1, KEK}_{\bar{K}N - \pi \Sigma}$
and $V^{\rm 2, KEK}_{\bar{K}N - \pi \Sigma}$ respectively.
Recently, new data on kaonic
hydrogen were obtained by the SIDDHARTA collaboration~\cite{SIDDHARTA1s}:
\begin{equation}
\label{SIDDHARTA}
\Delta E_{1s}^{\rm SIDD} = -283 \pm 36 \pm 6 \; {\rm eV}, \quad
\Gamma_{1s}^{\rm SIDD} = 541 \pm 89 \pm 22 \; {\rm eV}.
\end{equation}
Keeping this in mind, we obtained the new parameters for the
$\bar{K}N - \pi \Sigma$ potentials, used in~\cite{my_Kd} and
described in more details in~\cite{ourPRC_isobreak}.
The new one- and two-pole potentials reproducing
the experimental data~(\ref{SIDDHARTA}) will be denoted by
$V^{\rm 1, SIDD}_{\bar{K}N - \pi \Sigma}$ and
$V^{\rm 2, SIDD}_{\bar{K}N - \pi \Sigma}$ respectively.
In fact, the potentials reproducing the SIDDHARTA data reproduce
the KEK data as well since the $1 \sigma$ region of the most recent
experiment lies almost entirely inside the $1 \sigma$ KEK region.
\begin{center}
\begin{table}
\caption{Parameters of the new one- and two-pole $\bar{K}N - \pi \Sigma$
potentials: range $\beta^{\bar{\alpha}}$ $\left( {\rm fm}^{-1} \right)$,
strength $\lambda^{\bar{\alpha} \bar{\beta}}_I$ 
$\left( {\rm fm}^{-2} \right)$ parameters, and the additional
unitless parameter $s$ of the two-pole model.}
\label{params.tab}
\begin{center}
\begin{tabular}{ccc}
\hline \noalign{\smallskip}
{} & $V^{1,\rm SIDD}_{\bar{K}N - \pi \Sigma}$ &
     $V^{2,\rm SIDD}_{\bar{K}N - \pi \Sigma}$ \\
\noalign{\smallskip} \hline \noalign{\smallskip}
$\beta^{\bar{K}N}$ & $3.52$ & $3.64$ \\
$\beta^{\pi \Sigma}$ & $1.48$ & $1.05$ \\
$\lambda^{\bar{K}\bar{K}}_{0}$ & $-1.2777$ & $-1.3753$ \\
$\lambda^{\bar{K}\pi}_{0}$ & $0.5431$ & $0.5577$ \\
$\lambda^{\pi \pi}_{0}$ & $0.1541$ & $0.0518$ \\
$\lambda^{\bar{K}\bar{K}}_{1}$ & $0.4960 - i \, 0.1932$ 
  & $0.6178 - i \, 0.1965$ \\
$\lambda^{\bar{K}\pi}_{1}$ & $1.7085$ & $1.9151$ \\
$\lambda^{\pi \pi}_{1}$ & $1.9444$ & $1.9904$ \\
$s$ & $0.0000$ & $-0.7604$ \\
\noalign{\smallskip} \hline
\end{tabular}
\end{center}
\end{table}
\end{center}

The potential in momentum representation has the form
\begin{equation}
\label{Vseprb}
 V_{I}^{\bar{\alpha} \bar{\beta}}(k^{\bar{\alpha}},k'^{\bar{\beta}}) =
 \lambda_{I}^{\bar{\alpha} \bar{\beta}} \;
 g^{\bar{\alpha}}(k^{\bar{\alpha}}) \, g^{\bar{\beta}}(k'^{\bar{\beta}}),
\end{equation}
where indices $\bar{\alpha}, \bar{\beta} = 1, 2$ denote $\bar{K}N$
or $\pi \Sigma$ channel respectively, $I$ is a two-body isospin.
We used Yamaguchi form-factors in both channels for the one-pole potential 
$V^{\rm 1, SIDD}_{\bar{K}N - \pi \Sigma}$ and
in the $\bar{K}N$ channel for the two-pole potential
$V^{\rm 2, SIDD}_{\bar{K}N - \pi \Sigma}$:
\begin{equation}
\label{1res_ff}
 g^{\bar{\alpha}}(k^{\bar{\alpha}}) = \frac{1}{(k^{\bar{\alpha}})^2 +
 (\beta^{\bar{\alpha}})^2} \,,
\end{equation}
while in the $\pi \Sigma$ channel for the two-pole
$V^{\rm 2, SIDD}_{\bar{K}N - \pi \Sigma}$ we used the following form-factor
\begin{equation}
\label{2res_ffpi}
 g^{\bar{\alpha}}(k^{\bar{\alpha}}) = 
 \frac{1}{(k^{\bar{\alpha}})^2
   + (\beta^{\bar{\alpha}})^2} \,+\,
 \frac{s \, (\beta^{\bar{\alpha}})^2}{[(k^{\bar{\alpha}})^2 + 
 (\beta^{\bar{\alpha}})^2]^2} \,.
\end{equation}
Therefore, the two-pole potential contains additional to the strength
$\lambda_{I}^{\bar{\alpha} \bar{\beta}}$ and range $\beta^{\bar{\alpha}}_{I}$ 
parameters unitless parameter $s$.

The parameters of the new $V^{\rm 1, SIDD}_{\bar{K}N - \pi \Sigma}$ and 
$V^{\rm 2, SIDD}_{\bar{K}N - \pi \Sigma}$ potentials,
as before, were obtained by fitting to experimental data
on low-energy $K^- p$ scattering and kaonic hydrogen.
They are shown in Table~\ref{params.tab}. Both potentials reproduce the
medium values of the threshold branching ratios $\gamma$ and
$R_{\pi \Sigma}$, where
\begin{equation}
\label{gamma}
\gamma =
 \frac{\Gamma(K^- p \to \pi^+ \Sigma^-)}{\Gamma(K^- p \to
    \pi^- \Sigma^+)} = 2.36 \pm 0.04  \,,
\end{equation}
is the measured value~\cite{gammaKp1,gammaKp2} and
\begin{equation}
\label{RpiSigma}
R_{\pi \Sigma} =
 \frac{\Gamma(K^- p \to \pi^+ \Sigma^-)+\Gamma(K^- p \to \pi^- \Sigma^+)}{
 \Gamma(K^- p \to \pi^+ \Sigma^-) + \Gamma(K^- p \to \pi^- \Sigma^+) +
                                   \Gamma(K^- p \to \pi^0 \Sigma^0) } 
\end{equation}
is the ratio, constructed from the measured $R_c$ and $R_n$~\cite{gammaKp1,gammaKp2}:
\begin{eqnarray}
\label{Rc}
R_c &=& \frac{\Gamma(K^- p \to \pi^+ \Sigma^-, \pi^- \Sigma^+)}{\Gamma(K^- p \to
\mbox{all inelastic channels} )} = 0.664 \pm 0.011, \\
\label{Rn}
R_n &=& \frac{\Gamma(K^- p \to \pi^0 \Lambda)}{\Gamma(K^- p \to
\mbox{neutral states} )} = 0.189 \pm 0.015 \,.
\end{eqnarray}
An ``experimental'' value for the $R_{\pi \Sigma}$ is
\begin{equation}
 R_{\pi \Sigma} =  \frac{R_c}{1-R_n \, (1 - R_c)} \, = \, 0.709 \pm 0.011 \,.
\end{equation}
The elastic and inelastic $K^- p$ cross-sections $K^- p \to {K}^- p$,
$K^- p \to \bar{K}^0 n$, $K^- p \to \pi^+ \Sigma^-$, $K^- p \to \pi^- \Sigma^+$,
and $K^- p \to \pi^0 \Sigma^0$ are reproduced by the new potentials with
the same accuracy as in the previous article (see Fig.7 of~\cite{my_Kd}).

The remaining characteristics of the $V^{\rm 1, SIDD}_{\bar{K}N - \pi \Sigma}$
and  $V^{\rm 2, SIDD}_{\bar{K}N - \pi \Sigma}$ potentials
are shown in Table~\ref{phys_char.tab}. In particular, the ``strong''
pole positions $z_1$ and $z_2$ (in MeV) are presented there together with
the $1s$ level shifts $\Delta E_{1s}^{K^- p}$ and widths $\Gamma_{1s}^{K^- p}$
(eV) of kaonic hydrogen. The strong $K^- p$ scattering lengths, obtained from the
Lippmann-Schwinger coupled-channel equations and, therefore, 
exactly corresponding to the $\Delta E_{1s}^{K^- p}$ and $\Gamma_{1s}^{K^- p}$
values, are shown as well. The same characteristic of the
$V^{\rm 1, KEK}_{\bar{K}N - \pi \Sigma}$ and 
$V^{\rm 2, KEK}_{\bar{K}N - \pi \Sigma}$ potentials,
taken from Table II of \cite{my_Kd}, are also presented in the table.
It is useful for investigation of the role of the poles of the $\Lambda(1405)$
resonance in the three-body results. It is seen from Table~\ref{phys_char.tab}
that $V^{\rm 1, SIDD}_{\bar{K}N - \pi \Sigma}$
and  $V^{\rm 2, SIDD}_{\bar{K}N - \pi \Sigma}$ potentials with
different $z_1$ values and very close $\Delta E_{1s}^{K^- p}$,
$\Gamma_{1s}^{K^- p}$ together with $V^{\rm 1, KEK}_{\bar{K}N - \pi \Sigma}$
and  $V^{\rm 2, KEK}_{\bar{K}N - \pi \Sigma}$ with equal $z_1$
and different $\Delta E_{1s}^{K^- p}$, $\Gamma_{1s}^{K^- p}$
supplement each other.

\begin{figure}
\centering
\includegraphics[width=0.35\textwidth, angle=-90]{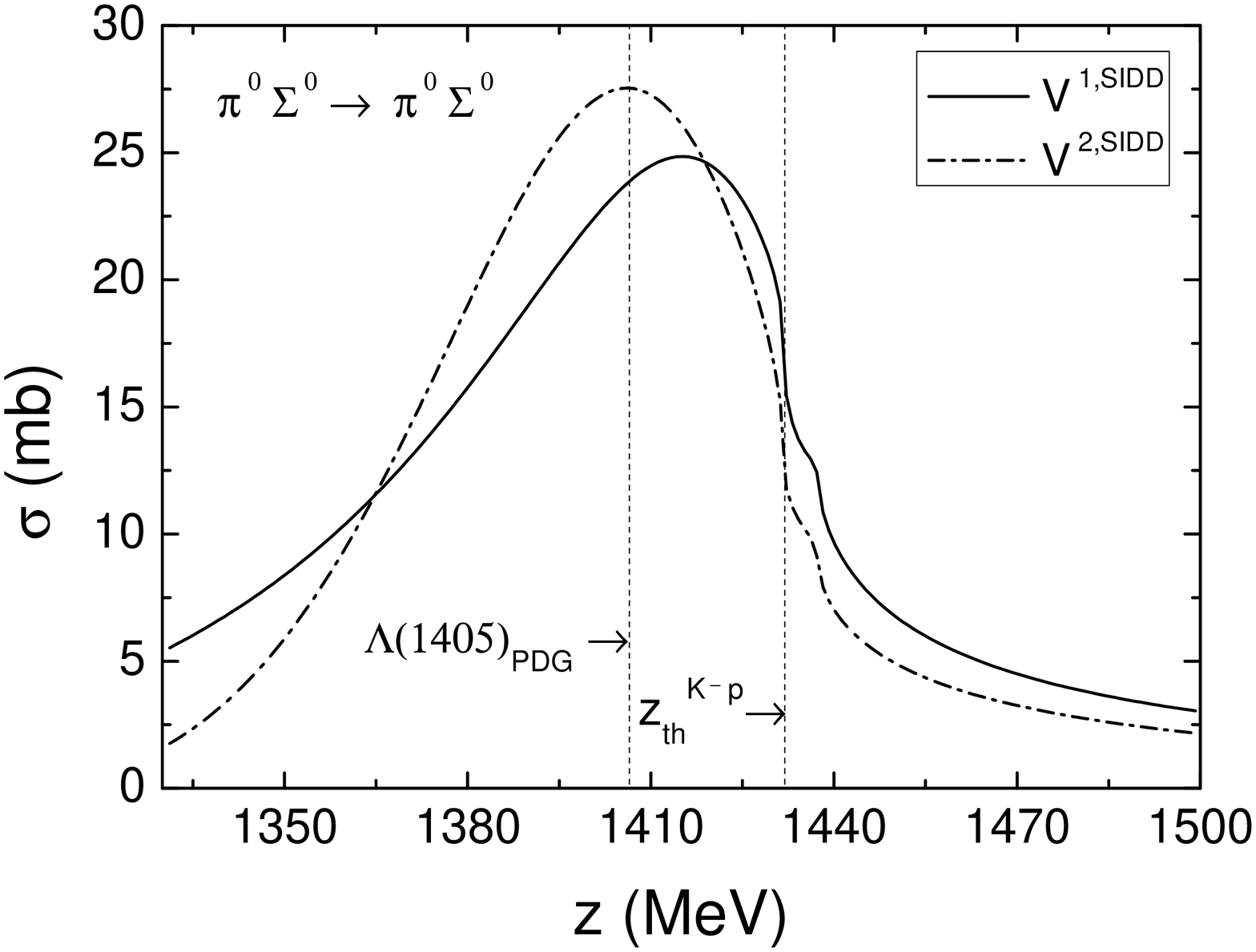}
\caption{Elastic $\pi^0 \Sigma^0$ cross-sections, obtained with
$V^{\rm 1, SIDD}_{\bar{K}N - \pi \Sigma}$ (solid line)
and  $V^{\rm 2, SIDD}_{\bar{K}N - \pi \Sigma}$ (dash-dotted line)
potentials. PDG value of the mass of $\Lambda(1405)$ resonance and
$K^- p$ threshold are shown as well (vertical lines).
\label{pi0Sig0.fig}}
\end{figure}

The parameters of the highest strong pole $z_1$  for 
$V^{\rm 1, SIDD}_{\bar{K}N - \pi \Sigma}$
and  $V^{\rm 2, SIDD}_{\bar{K}N - \pi \Sigma}$
differ significantly
from the PDG values of the mass and width of the $\Lambda(1405)$
resonance~\cite{PDG}:
\begin{equation}
\label{MG_PDG}
M^{PDG}_{\Lambda(1405)} = 1406.5 \pm 4.0 \; {\rm MeV}, \quad
\Gamma^{PDG}_{\Lambda(1405)} = 50 \pm 2.0 \; {\rm MeV}.
\end{equation}
However, the elastic $\pi^0 \Sigma^0$ cross-sections obtained with the
potentials have peaks with the maxima, which almost coincide with
the $M^{PDG}_{\Lambda(1405)}$ value
(for $\left. V^{\rm 2, SIDD}_{\bar{K}N - \pi \Sigma} \right)$ or is
closer to it than $Re(z_1)$ 
$\left( V^{\rm 1, SIDD}_{\bar{K}N - \pi \Sigma} \right)$,
see Fig.~\ref{pi0Sig0.fig}. The widths of the peaks
are within experimental errors of $\Gamma^{PDG}_{\Lambda(1405)}$.

In the three-body equations we used isospin-averaged masses
for the particles instead of the physical ones.
Inserted into the Lippmann-Schwinger equations with isospin-averaged
masses the $\bar{K}N - \pi \Sigma$ potentials with parameters shown
in Table~\ref{params.tab}
give $K^- p$ scattering lengths $a_{K^- p}^{\rm aver}$, which are
different from the ``physical'' $a_{K^- p}$ values. We show them
in Table~\ref{phys_char.tab} together
with $a_{\bar{K}N,0}^{\rm aver}$ and $a_{\bar{K}N,1}^{\rm aver}$,
which are the $\bar{K}N$ scattering lengths for isospin $I=0$ and $I=1$
states respectively. In the case of the isospin-averaged masses
$a_{\bar{K}N,1}^{\rm aver}$ coincides with the $K^- n$ scattering
length. 

The pole positions $z_1$ and $z_2$ remain almost unchanged,
the $K^- p$ ``averaged'' cross-sections do not manifest
$\bar{K}^0 n$ threshold, but still describes the experimental
data very well (see Fig.~7 of \cite{my_Kd}).

\subsection{Three-body formalism and the rest of the two-body input}
\label{formalism.sect}

The near-threshold elastic $K^- d$ amplitudes were obtained
using the Fad\-deev-type Alt-Grassberger-Sandhas (AGS)
equations with coupled $\bar{K}NN$
and $\pi \Sigma N$ channels, described in details in~\cite{my_Kd}.
The total energy of the $K^- d$ system was considered 
up to the three-body $\bar{K}NN$ breakup threshold,
where kinetic energy of the relative $K^- - d$ motion $z_{kin}$ is
equal to the deuteron binding energy $E_{\rm deu}$.

The spin of the $K^- d$ system is equal to one,
the isospin to one half. All two-body potentials
as well as the corresponding to them two-body $T$-matrices, 
being an input for the three-body calculations, were chosen to
have zero orbital momentum. The total orbital momentum was also chosen
equal to zero, therefore, the total angular momentum
of the three-body system is equal to one. 
The calculations were performed in the momentum representation
with proper antisymmetrization, necessary due to presence of 
the two baryons in every channel.
The obtained system of ten integral equations was solved
numerically, the logarithmic singularities in the $\pi \Sigma N$
channel and a pole at the deuteron binding energy were properly
taken into account.

In contrast to the two-body coupled-channel equations for the
$\bar{K}N - \pi \Sigma$ potentials construction, we 
neither included Coulomb interaction nor used physical masses for 
the particles in the three-body equations. The effect of inclusion 
of the physical masses of $K^-$ and $\bar{K}^0$ into AGS equation
was studied recently in~\cite{Revai}. It makes $2\%$ difference from
the case of isospin-averaged masses for the real part of the
$K^- d$ scattering length, while the imaginary part remains almost
unchanged.

We used one or two-term separable potentials, which are
isospin and spin dependent. The new one- and two-pole
$\bar{K}N - \pi \Sigma$ potentials reproducing the SIDDHARTA data
on kaonic hydrogen are described
in the previous subsection. In spite of the fact that SIDDHARTA
experimental results~(\ref{SIDDHARTA}) are more accurate that
KEK data~(\ref{KEK}), we used the $V^{\rm 1, KEK}_{\bar{K}N - \pi \Sigma}$
and $V^{\rm 2, KEK}_{\bar{K}N - \pi \Sigma}$ potentials from
Ref.~\cite{my_Kd} as well for investigation of the role of the
poles of the $\Lambda(1405)$ resonance in the three-body results.
As for the rest of the potentials necessary for the calculations,
we used the best $NN$ and $\Sigma N$ potentials from
the previous calculation and, as before, neglected $\pi N$
interaction (due to its weakness in the $l=0$ state).

The two-term $NN$ potential $V_{NN}^{\rm TSA-B}$~\cite{DolesNN}
(Eqs.(13) and (19) of~\cite{my_Kd}), chosen for the present calculation,
accurately reproduces
Argonne V18 phase shifts and, therefore,
is repulsive at short distances. It gives correct scattering
length $a = -5.413$ fm, effective range $r_{\rm eff} = 1.760$ fm
of the ${}^3S_1$ $NN$ state and accurate binding energy of the
deuteron $E_{\rm deu}=2.2246$ MeV.
We used exact optical spin-dependent potential
$V_{\Sigma N}^{\rm Sdep,Opt}$ constructed in~\cite{my_Kd}
for description of the $\Sigma N (- \Lambda N)$ interaction.
The potential reproduces the experimental
data on $\Sigma N$ and $\Lambda N$ scattering quite well. 

\begin{center}
\begin{table}
\caption{Physical characteristics of the $V^{1,SIDD}_{\bar{K}N - \pi \Sigma}$ 
and $V^{2,SIDD}_{\bar{K}N - \pi \Sigma}$ potentials: 
strong pole positions $z_1$ and $z_2$ (MeV), $K^- p$ scattering length
$a_{K^- p}$ (fm), kaonic hydrogen $1s$ level shift $\Delta E_{1s}^{K^- p}$ and
width $\Gamma_{1s}^{K^- p}$ (eV). The same values for
the $V^{1,KEK}_{\bar{K}N - \pi \Sigma}$ 
and $V^{2,KEK}_{\bar{K}N - \pi \Sigma}$ potentials are taken from~\cite{my_Kd}.
Scattering lengths, calculated with isospin-averaged masses: $a_{K^- p}^{\rm aver}$,
$a_{\bar{K}N,0}^{\rm aver}$ and $a_{\bar{K}N,1}^{\rm aver}$ are also presented.
The three-body characteristics, obtained using the corresponding
$\bar{K}N - \pi \Sigma$ potentials, are shown as well:
$K^- d$ scattering length $a_{K^- d}$ (fm)
($a_{K^- d}$ values obtained using $V^{1,KEK}_{\bar{K}N - \pi \Sigma}$ 
and $V^{2,KEK}_{\bar{K}N - \pi \Sigma}$ potentials
are taken from ~\cite{my_Kd}), effective range $r^{\rm eff}_{K^- d}$ (fm),
and kaonic deuterium characteristics $\Delta E_{1s}^{K^- d}$ and width 
$\Gamma_{1s}^{K^- d}$ (eV).}
\label{phys_char.tab}
\begin{tabular}{ccccc}
\hline \noalign{\smallskip}
 & $V^{\rm 1,SIDD}_{\bar{K}N - \pi \Sigma}$ & $V^{\rm 2,SIDD}_{\bar{K}N - \pi \Sigma}$
 & $V^{\rm 1,KEK}_{\bar{K}N - \pi \Sigma}$ & $V^{\rm 2,KEK}_{\bar{K}N - \pi \Sigma}$ \\
\noalign{\smallskip} \hline \noalign{\smallskip}
$z_1$ & $1426 - i \, 48$  
      & $1414 - i \, 58$  
      & $1409 - i \, 36$
      & $1409 - i \, 36$\\
$z_2$ &  $-$   
      &  $1386 - i \, 104$ 
      & $-$
      & $1381 - i \, 105$\\
\noalign{\smallskip} \hline \noalign{\smallskip}
$a_{K^- p}$  & $-0.76 + i \, 0.89$ & $-0.74 + i \, 0.90$
			& $-1.00 + i \, 0.68$ & $-0.96 + i \, 0.80$ \\
$\Delta E_{1s}^{K^- p}$  & $-313$ &  $-308$
						& $-377$ &  $-373$ \\
$\Gamma_{1s}^{K^- p}$    & $597$ &  $602$
						& $434$ &  $514$ \\
\noalign{\smallskip} \hline \noalign{\smallskip}
$a_{K^- p}^{\rm aver}$  & $-0.52 + i \, 0.81$ & $-0.48 + i \, 0.82$
			& $-0.80 + i \, 0.62$ & $-0.72 + i \, 0.73$ \\
$a_{\bar{K}N,0}^{\rm aver}$  & $-1.45 + i \, 0.86$ & $-1.45 + i \, 0.86$
			& $-1.60 + i \, 0.67$ & $-1.62 + i \, 0.78$ \\
$a_{\bar{K}N,1}^{\rm aver}$  & $0.41 + i \, 0.75$ & $0.48 + i \, 0.77$
			& $-0.004 + i \, 0.56$ & $0.18 + i \, 0.67$ \\
\noalign{\smallskip} \hline \noalign{\smallskip}
$a_{K^- d}$  & $-1.48 + i \, 1.22$ & $-1.51 + i \, 1.23$
			& $-1.49 + i \, 0.98$ & $-1.57 + i \, 1.11$ \\
$r^{\rm eff}_{K^- d}$  & $0.68 - i \, 1.33$ & $0.67 - i \, 1.35$
			& $0.55 - i \, 1.15$ & $0.62 - i \, 1.19$ \\
$\Delta E_{1s}^{K^- d}$  & $-781$ &  $-794$
						& $-767$ &  $-809$ \\
$\Gamma_{1s}^{K^- d}$    & $1010$ &  $1012$
						& $810$ &  $902$ \\
\noalign{\smallskip} \hline
\end{tabular}
\end{table}
\end{center}

\subsection{Elastic $K^- d$ amplitudes: results}
\label{3bodyResults.sect}

\begin{figure}
\centering
\includegraphics[width=0.35\textwidth, angle=-90]{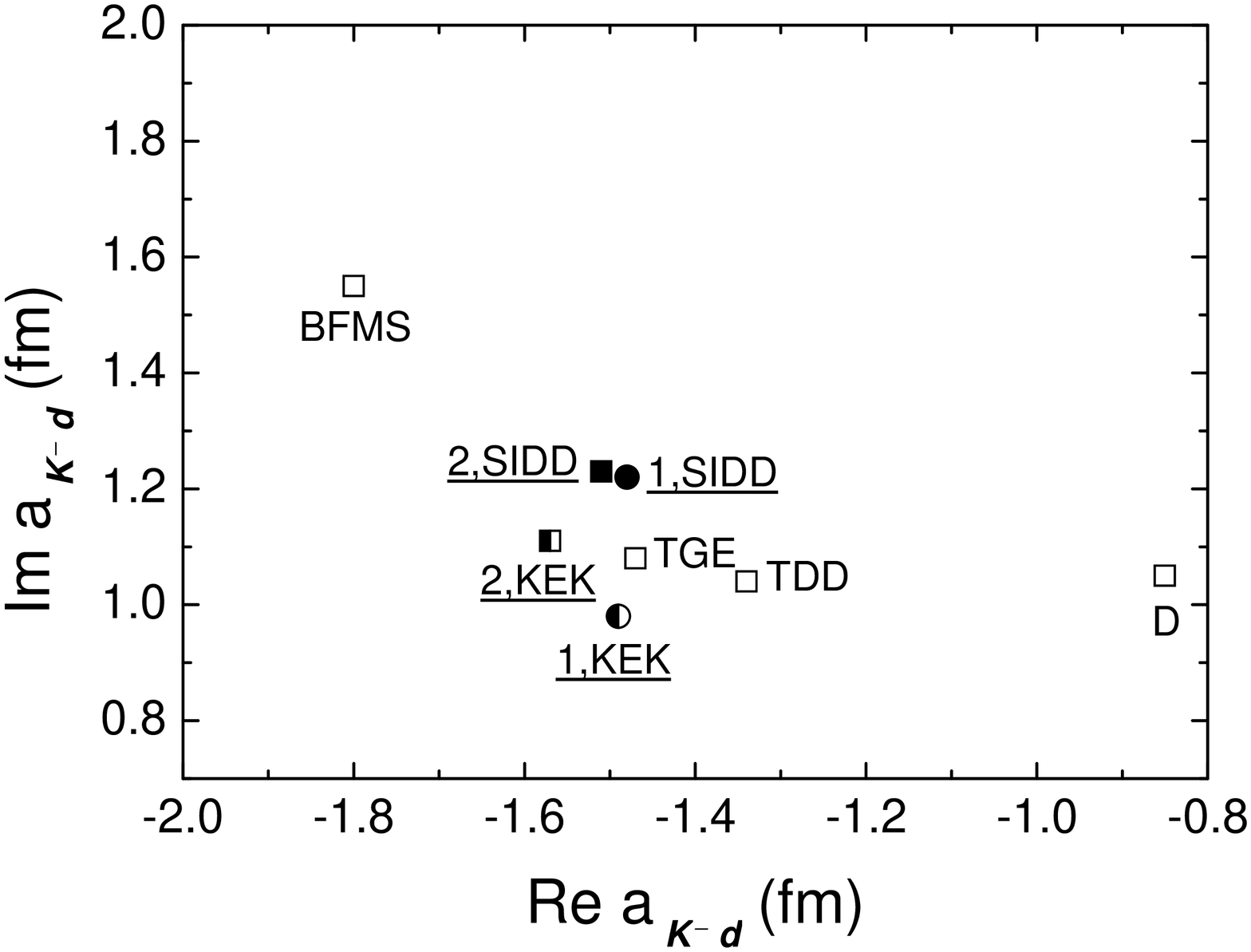}
\caption{The results of the $K^- d$ scattering length calculations
using $V^{\rm 1, SIDD}_{\bar{K}N - \pi \Sigma}$ (black circle)
and  $V^{\rm 2, SIDD}_{\bar{K}N - \pi \Sigma}$ (black square)
potentials. Results of our previous calculations~\cite{my_Kd}
with one- (half-empty circle) and two-pole (half-empty square)
representative potentials reproducing KEK data on kaonic hydrogen
are shown as well together with earlier $a_{K^- d}$ results:
BFMS~\cite{Kd_BFMS}, TGE~\cite{Kd_TGE}, TDD~\cite{Kd_TDD},
D~\cite{Kd_Deloff} (empty squares).
\label{aKd.fig}}
\end{figure}

The $K^- d$ scattering lengths obtained using
the new one- and two-pole $\bar{K}N - \pi \Sigma$ potentials,
reproducing the SIDDHARTA data on kaonic hydrogen,
are shown in Fig.~\ref{aKd.fig}  (black circle for $a^{\rm 1, SIDD}_{K^- d}$ 
and black square for $a^{\rm 2, SIDD}_{K^- d}$)
together with our representative $a_{K^- d}$ values from~\cite{my_Kd},
obtained using the $V^{\rm 1, KEK}_{\bar{K}N - \pi \Sigma}$ and 
$V^{\rm 2, KEK}_{\bar{K}N - \pi \Sigma}$ potentials (half-empty
circle and half-empty square respectively). They are also
shown in Table~\ref{phys_char.tab}, where we collect two-body
and three-body characteristics of all four our $\bar{K}N - \pi \Sigma$
potentials.

It is seen, that the imaginary parts of the new $a_{K^- d}^{\rm SIDD}$
are larger that those of the previous $a_{K^- d}^{\rm KEK}$ by
$20\%$ for the one-pole and by $10\%$ for the two-pole versions.
The real parts are not so different, especially for the one-pole
potentials. However, the main difference of the new results from
those of~\cite{my_Kd} is that the $K^- d$ scattering
lengths obtained with the one- and two-pole $\bar{K}N - \pi \Sigma$
potentials reproducing the SIDDHARTA data on kaonic hydrogen 
are very close one to the other, while our $a_{K^- d}^{\rm KEK}$ values
obtained with the one- and two-pole potentials are rather different.
Therefore, it is not possible to resolve the question of the number
of the poles forming the $\Lambda(1405)$ resonance from the results
on near-threshold elastic $K^- d$ scattering.

Four more values of the $K^- d$ scattering length
from Refs.~\cite{Kd_BFMS,Kd_TGE,Kd_TDD,Kd_Deloff} (empty squares),
where the Faddeev equations were also used, are shown in the
Fig.~\ref{aKd.fig} as well. Three of them~\cite{Kd_BFMS,Kd_TGE,Kd_TDD}
were obtained with coupled channels, while the fourth
one~\cite{Kd_Deloff} is a result of a single-channel calculation.
The four $a_{K^- d}$ values were compared with our previous results 
in~\cite{my_Kd}, and since our present $a_{K^- d}$ values are not too
far from the previous ones, it is not necessary to repeat
the discussion. We do not show results obtained
using Fixed Scatterer Approximation since it was shown
in~\cite{my_Kd}, that the method is not a proper approach
for the $K^- d$ system.

\begin{figure}
\centering
\includegraphics[width=0.35\textwidth, angle=-90]{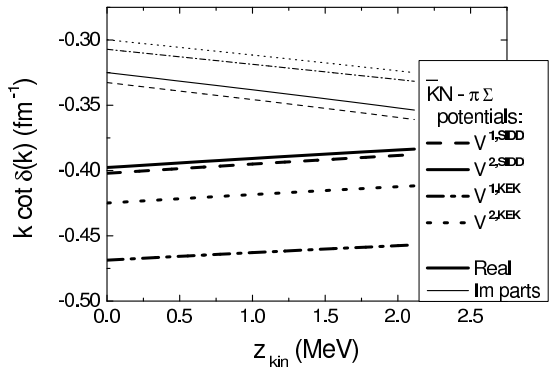}
\caption{Real (thick lines) and imaginary (thin lines) parts of
the elastic near-threshold $K^- d$ amplitudes presented in a form of
$k \cot \delta(k)$ function. The results obtained from the
coupled-channel three-body AGS equations using 
$V^{\rm 1,SIDD}_{\bar{K}N - \pi \Sigma}$ (dashed lines),
$V^{\rm 2,SIDD}_{\bar{K}N - \pi \Sigma}$ (solid lines),
$V^{\rm 1,KEK}_{\bar{K}N - \pi \Sigma}$ (dotted lines) and
$V^{\rm 2,KEK}_{\bar{K}N - \pi \Sigma}$ (dash-dotted lines)
potentials are shown.
\label{amplitudes.fig}}
\end{figure}

The elastic amplitudes of $K^- d$ scattering
for kinetic energy from zero to $E_{\rm deu}$ calculated using
four versions of $\bar{K}N - \pi \Sigma$ potentials are shown in
Fig.~\ref{amplitudes.fig}. They are presented in a form of
$k \cot \delta(k)$ function.
Real (thick lines) and imaginary (thin lines)
parts of the function obtained with 
$V^{\rm 1,SIDD}_{\bar{K}N - \pi \Sigma}$ (dashed lines),
$V^{\rm 2,SIDD}_{\bar{K}N - \pi \Sigma}$ (solid lines),
$V^{\rm 1,KEK}_{\bar{K}N - \pi \Sigma}$ (dotted lines) and
$V^{\rm 1,KEK}_{\bar{K}N - \pi \Sigma}$ (dash-dotted lines)
potentials are plotted. The chosen representation of the results
demonstrates that the elastic near-threshold $K^- d$ amplitudes
can be approximated by the effective range expansion rather
accurately since the lines are almost straight. 
The calculated effective ranges $r^{\rm eff}_{K^- d}$ of
$K^- d$ scattering are shown in Table~\ref{phys_char.tab}.

\section{Kaonic deuterium}
\label{deuterium.sec}

Mixture of the strong and Coulomb interaction makes
investigation of low levels of hadronic atoms with the help of
the Faddeev equations difficult. Due to this we constructed a complex
two-body $K^- - d$ potential and used it for investigation of the $1s$ level
of kaonic deuterium by Lippmann-Schwinger equation.

A one-term separable potential is too simple for reproducing the
$K^- d$ amplitudes obtained using AGS equations, therefore we chose
the two-term form:
\begin{equation}
\label{VKd}
V_{K^- d}(\vec{k},\vec{k}') = \lambda_{1,K^- d} \, g_1(\vec{k}) g_1(\vec{k}')
+ \lambda_{2,K^- d} \, g_2(\vec{k}) g_2(\vec{k}')
\end{equation}
with Yamaguchi form-factors
\begin{equation}
\label{gKd}
g_i(k) = \frac{1}{\beta_{i,K^- d}^2 + k^2}, \qquad i=1,2.
\end{equation}

The complex strength parameters $\lambda_{1,K^- d}$ and $\lambda_{2,K^- d}$
were fixed by the conditions, that the $V_{K^- d}$ potential reproduces
the $K^- d$ scattering length $a_{K^- d}$ and the effective range
$r^{\rm eff}_{K^- d}$ obtained with one of the four $\bar{K}N - \pi \Sigma$
potentials and shown in Table~\ref{phys_char.tab}. Variation of the real
$\beta_{1,K^- d}$ and $\beta_{2,K^- d}$ parameters allowed us to 
reproduce the full near-threshold amplitudes more accurately. As
a result, the near-threshold amplitudes obtained from the three-body
calculations $f^{(3)}_{K^- d}$, are reproduced by the two-body
$K^- - d$ potentials through the interval $[0,E_{\rm deu}]$ with such
accuracy, that the two-body functions $k \cot \delta^{(2)}(k)$
are indistinguishable from the three-body $k \cot \delta^{(3)}(k)$
in Fig.\ref{amplitudes.fig}. 

We denote $K^- - d$ potentials, which reproduce three-body amplitudes
obtained using $V^{\rm 1,SIDD}_{\bar{K}N - \pi \Sigma}$ and
$V^{\rm 2,SIDD}_{\bar{K}N - \pi \Sigma}$ potentials, as
$V^{\rm 1,SIDD}_{K^- d}$ and $V^{\rm 2,SIDD}_{K^- d}$ respectively
($V^{\rm 1,KEK}_{K^- d}$ and $V^{\rm 2,KEK}_{K^- d}$ are defined
analogically).
The parameters of the potentials are shown in Table~\ref{paramsKd.tab}.
As is seen, both $\beta_{1,K^- d}$ and
$\beta_{2,K^- d}$ parameters for every potential turned up to be much
smaller then the corresponding $\beta^{\bar{K}N}$ parameter,
as it can be expected having in mind the size of the deuteron.
\begin{center}
\begin{table}
\caption{Range $\beta_{i,K^- d}$ $\left( {\rm fm}^{-1} \right)$
and strength $\lambda_{i,K^- d}$ $\left( {\rm fm}^{-2} \right)$
parameters of the two-term $V^{\rm 1,SIDD}_{K^- d}$ and
$V^{\rm 2,SIDD}_{K^- d}$ potentials ($i=1,2$).
}
\label{paramsKd.tab}
\begin{center}
\begin{tabular}{ccc}
\hline \noalign{\smallskip}
{} & $V^{1,\rm SIDD}_{K^- d}$ &
     $V^{2,\rm SIDD}_{K^- d}$ \\
\noalign{\smallskip} \hline \noalign{\smallskip}
$\beta_{1,K^- d}$ & $0.7$ & $0.7$ \\
$\beta_{2,K^- d}$ & $1.0$ & $1.0$ \\
$\lambda_{1,K^- d}$ & $-0.0051 - i\, 0.0012$ & $-0.0056 - i\, 0.0016$ \\
$\lambda_{2,K^- d}$ & $0.0096 - i\, 0.0420$ & $0.0109 - i\, 0.0420$ \\
\noalign{\smallskip} \hline
\end{tabular}
\end{center}
\end{table}
\end{center}

The constructed two-body complex potentials $V_{K^- d}$ were used
for calculation of the $1s$ atomic level of kaonic deuterium, we
are interested in. 
Calculation of the binding energy of a two-body system,
described by the Hamiltonian with strong $V^S$ and Coulomb $V^{Coul}$
interactions
\begin{equation}
H = H^0 + V^S + V^{Coul}
\end{equation}
was performed, for example, in~\cite{ourPRC_isobreak}, where 
the one-term $V^S$ describing interaction of
the coupled $\bar{K}N$ and $\pi \Sigma$ channels
was considered. For the two-term one-channel separable potential
$V^S = V_{K^- d}$, defined by Eq.(\ref{VKd}),
the binding energy condition can be written explicitly:
\begin{eqnarray}
\nonumber
\left[ \lambda_{1,K^- d}^{-1} - 
\langle g_1 | G^{Coul}(E_{1s}^{S+Coul}) | g_1 \rangle \right] \,
\left[ \lambda_{2,K^- d}^{-1} - 
\langle g_2 | G^{Coul}(E_{1s}^{S+Coul}) | g_2 \rangle \right] &{}&\\
\label{SCeq}
- \langle g_1 | G^{Coul}(E_{1s}^{S+Coul}) | g_2 \rangle \,
\langle g_2 | G^{Coul}(E_{1s}^{S+Coul}) | g_1 \rangle
= 0, &{}&
\end{eqnarray}
where $G^{Coul}$ is the Coulomb Green's function
\begin{equation}
G^{Coul}(z) = (z - H^0 - V^{Coul})^{-1}.
\end{equation}
The expression for the matrix element of the $G^{Coul}$ function
with the Yamaguchi form-factors $| g \rangle$,
entering Eq.(\ref{SCeq}), can be found in~\cite{gGg_Coul3}. 

The real part of the binding energy of the $1s$ level of the kaonic
deuterium, obtained from Eq.~(\ref{SCeq}), defines
the $1s$ level shift:
\begin{equation}
\Delta E_{1s} = E^{Coul}_{1s} -
Re \left( E^{S+Coul}_{1s} \right) \,,
\end{equation}
where $E^{Coul}_{1s}$ is the pure Coulomb energy. 
The doubled imaginary part of the binding
energy $E^{S+Coul}_{1s}$ defines the width of the ground state
$\Gamma_{1s}$.

\section{Results}
\label{results.sec}

\begin{figure}
\centering
\includegraphics[width=0.35\textwidth, angle=-90]{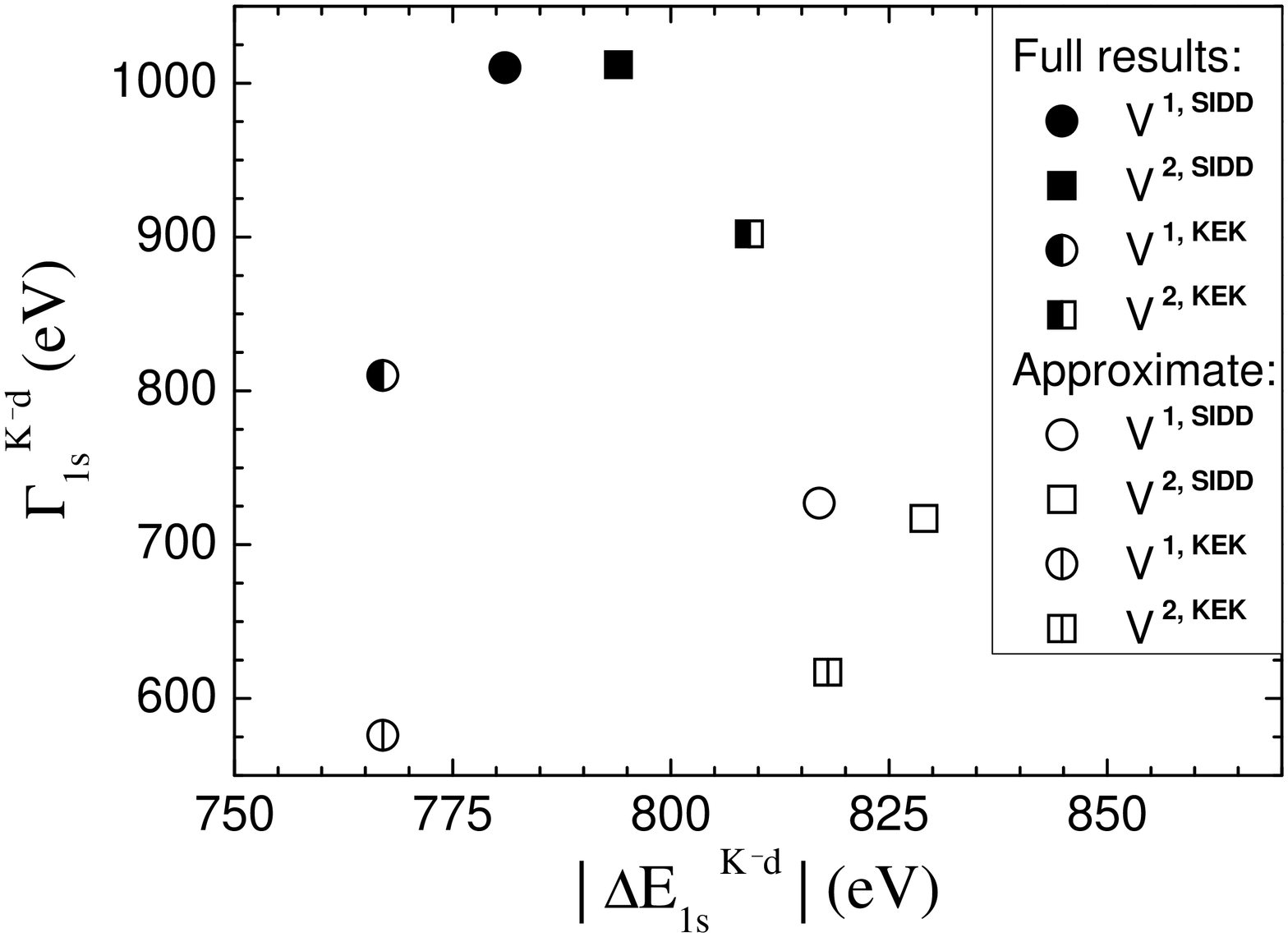}
\caption{Level shifts and widths of kaonic deuterium, calculated
with $V^{\rm 1,SIDD}$ (black circle), $V^{\rm 2,SIDD}$ (black square).
$V^{\rm 1,KEK}$ (black-and-white circle) and $V^{\rm 2,KEK}$ 
(black-and-white square) potentials. Approximate values
obtained using corrected Deser formula are also shown
(white circle, white square, crossed circle and crossed square,
respectively).
\label{Kd_results}}
\end{figure}

The level shifts $\Delta E_{1s}^{K^- d}$ and widths $\Gamma_{1s}^{K^- d}$
of kaonic deuterium obtained with the two-body $V_{K^- d}$ potentials
are shown in Fig.~\ref{Kd_results}. The characteristics of the $1s$ level
of kaonic deuterium calculated using the four versions of
the $K^- - d$ potential, are plotted in:
black circle $\left( V^{\rm 1,SIDD}_{K^- d} \right)$,
black square $\left( V^{\rm 2,SIDD}_{K^- d} \right)$,
black-and-white circle $\left( V^{\rm 1,KEK}_{K^- d} \right)$
black-and-white square $\left( V^{\rm 2,KEK}_{K^- d} \right)$.
The results are also shown in Table~\ref{phys_char.tab}.

The $1s$ level characteristics of kaonic deuterium
corresponding to the $V^{\rm 1,KEK}_{\bar{K}N - \pi \Sigma}$
and $V^{\rm 2,KEK}_{\bar{K}N - \pi \Sigma}$  potentials
with one- and two-pole structure of the $\Lambda(1405)$ resonance
respectively, differ by $5\%$ in the level shifts and by $10\%$ in
the widths. It could give possibility to draw conclusions about nature
of the $\Lambda(1405)$ resonance from comparison with some
(precise enough) experimental data. However, the results corresponding
to the $V^{\rm 1,SIDD}_{\bar{K}N - \pi \Sigma}$ and
$V^{\rm 2,SIDD}_{\bar{K}N - \pi \Sigma}$ potentials have almost 
the same widths and very close shifts.
Therefore, an experiment on the kaonic
deuterium characteristics as well as the $K^- d$ scattering data
cannot give an answer to the question on
the number of the poles forming the $\Lambda(1405)$ resonance.
Moreover, rather large widths, especially those obtained with
the new $\bar{K}N - \pi \Sigma$ potentials, rise the question whether
it will be possible to measure these characteristics accurately.

Figure~\ref{Kd_results} also contains the results obtained using an
approximate relation. A ``corrected Deser'' formula
(Eq.~(19) of \cite{corDeser}),
connecting a scattering length with characteristics
of a hadronic atom and, according to the
authors, containing isospin-breaking corrections
in respect to the original Deser formula~\cite{Deser},
is widely used nowadays. 
It was shown in~\cite{ourPRC_isobreak,cieply}, that the accuracy
of the formula is about $10\%$
for the two-body $K^- p$ system. Obviously, the accuracy of the
approximate formula should be less for the three-body $K^- d$ system
since three-body effects cannot be taken into account in it.

The $1s$ level shifts and widths of kaonic deuterium
obtained from the corrected Deser formula using
four our $a_{K^- d}$
values are shown in Fig.~\ref{Kd_results} in the empty circle
$\left( a^{\rm 1,SIDD}_{K^- d} \right.$ is used), empty square
$\left( a^{\rm 2,SIDD}_{K^- d} \right)$,
vertically crossed circle $\left( a^{\rm 1,KEK}_{K^- d} \right)$
and vertically crossed square $\left( a^{\rm 2,KEK}_{K^- d} \right)$.
The top index of $a_{K^- d}$ denotes the $\bar{K}N - \pi \Sigma$
potential, which was used in the AGS equations.
It is seen, that the level shifts obtained using the approximate
formula are very close to ours for the KEK and have $4\%$ difference
for the SIDDHARTA $K^- d$ scattering lengths.
As for the width of the $1s$ atom level,
the approximate formula underestimates it at about $30\%$
in comparison with all four our results, calculated using
the two-body $V_{K^- d}$ potentials.

Now, in contrast to the kaonic hydrogen case, our results of
the characteristics of kaonic deuterium cannot be
called ``exact''. However, the only shortcoming of our
approach, which is the supposed structurelessness of the deuteron,
is contained in the corrected Deser formula as well. Indeed,
the only change, which can be made in the original formula
for the $K^- p$ system for it's usage in the $K^- d$ case,
is the change of the corresponding reduced mass.
Thus, the deuteron structure is not taken into account in the
approximate formula. In contrast to~\cite{corDeser}, we made
no additional approximations, therefore,
our approach is much more accurate.

Table~\ref{phys_char.tab} containing several two-body characteristics
of the four $\bar{K}N - \pi \Sigma$ potentials
and the corresponding to them three-body $K^- d$ scattering
and kaonic deuterium observables allows us to reveal the role
of the $\bar{K}N - \pi \Sigma$ interaction.
Indeed, the three-body formalisms together with the two-body input,
except the main interaction, are the same in all
three-body calculations. Therefore, differences between $a_{K^- d}$,
$r^{\rm eff}_{K^- d}$, $\Delta E_{1s}^{K^- d}$ and
$\Gamma_{1s}^{K^- d}$ are caused by differences of
the $V^{\rm 1, KEK}_{\bar{K}N - \pi \Sigma}$,
$V^{\rm 2, KEK}_{\bar{K}N - \pi \Sigma}$,
$V^{\rm 1, SIDD}_{\bar{K}N - \pi \Sigma}$
and  $V^{\rm 2, SIDD}_{\bar{K}N - \pi \Sigma}$ potentials.

Data in Table~\ref{phys_char.tab} confirms, that the $K^- d$ scattering
lengths have no correlation with $z_1$ (and $z_2$)
pole properties at all: equal $z_1$ values for the two potentials
reproducing the KEK data lead to sufficiently different
$a^{1,KEK}_{K^- d}$ and $a^{2,KEK}_{K^- d}$ values.
And conversely, very close $K^- d$ scattering lengths
were obtained with
$V^{\rm 1, SIDD}_{\bar{K}N - \pi \Sigma}$
and  $V^{\rm 2, SIDD}_{\bar{K}N - \pi \Sigma}$, which are
characterised by rather different $z_1$ values.
The same is true for the effective range $r^{\rm eff}_{K^- d}$, 
level shift $\Delta E_{1s}^{K^- d}$ and width $\Gamma_{1s}^{K^- d}$
of kaonic deuterium. Therefore, there is no correlation between
the pole or poles of the $\Lambda(1405)$ resonance described
by a $\bar{K}N - \pi \Sigma$ potential and the three-body
$K^- d$ elastic scattering or kaonic deuterium characteristic
obtained using the potentials.

On the other hand, there is a clear correlation between the
imaginary parts of the $K^- d$ scattering length $a_{K^- d}$
and the width of kaonic deuterium $\Gamma_{1s}^{K^- d}$ --
the same as between the two-body Im $a_{K^- p}$ and
$\Gamma_{1s}^{K^- p}$. Such property was used in the original
Deser formula~\cite{Deser} already, but the dependence definitely
differs from the original Deser as well as the corrected Deser
functional form for the $K^- p$ and $K^- d$ systems.

\section{Conclusions}
\label{conclusions.sec}

We calculated $1s$ level shift $\Delta E_{1s}^{K^- d}$
and width $\Gamma_{1s}^{K^- d}$ of kaonic deuterium,
corresponding to the new $\bar{K}N - \pi \Sigma$ potentials,
reproducing SIDDHARTA data on kaonic hydrogen. Our previous potentials,
reproducing KEK data, were used as well. The results
were obtained from the Lippmann-Schwinger equation with directly
included Coulomb and the strong two-body $K^- - d$ potential.
The $K^- - d$ potentials reproduce the three-body characteristics:
$a_{K^- d}$, $r^{\rm eff}_{K^- d}$ and the elastic near-threshold $K^- d$
amplitudes, which were calculated using the coupled-channel AGS equations.
All results are shown in Table~\ref{phys_char.tab} together with
several characteristics of the four $\bar{K}N - \pi \Sigma$ potentials.

We also checked, whether it is possible to resolve the question
of the number of poles forming $\Lambda(1405)$ resonance
from comparison of theoretical
predictions with eventual experimental data on low-energy $K^- d$
scattering or kaonic deuterium. For this purpose the one- and two-pole
versions of the $\bar{K}N - \pi \Sigma$ potentials were used.
We found, that the question cannot be resolved in such a way
since the results, corresponding to the new one- and two-pole
$\bar{K}N - \pi \Sigma$ potentials are very close. Moreover,
it is seen from Table~\ref{phys_char.tab} that there is no correlation
between the pole or poles of the $\Lambda(1405)$ resonance described
by a $\bar{K}N - \pi \Sigma$ potential and the three-body
$K^- d$ elastic scattering or kaonic deuterium characteristic
obtained using the potential.

The calculated  widths of the $1s$ level of kaonic deuterium
are rather large, thus the question of possibility of accurate
measuring the atomic characteristics rises. 
We demonstrated, that the corrected Deser formula, widely used
for estimation of the $1s$ level shift and width of kaonic atoms,
considerably underestimates the width of kaonic deuterium.

\section*{Acknowledgments}
The author is thankful to J. R\'evai for fruitful discussions.
The work was supported by the GACR grant P202/12/2126.

\end{document}